\documentclass[
showpacs,
floatfix,
aps,
prl,
amsmath,
twocolumn,
superscriptaddress,
tightenlines
]{revtex4}

\usepackage{graphicx}
\usepackage{dcolumn}
\usepackage{amsmath}
\usepackage{bm}
\usepackage{times}
\usepackage[varg]{txfonts}
\def\eac{\epsilon^{\mbox{{\scriptsize ac}}}}
\def\edc{\epsilon^{\mbox{\scriptsize dc}}}
\def\eph{\epsilon^{\mbox{\scriptsize ph}}}
\def\pac{\phi}

\def\oc{\omega_{\mbox{\scriptsize {C}}}}
\def\oh{\omega_{\mbox{\scriptsize {H}}}}
\def\os{\omega_s}
\def\olo{\omega_{\mbox{\scriptsize {LO}}}}
\def\rc{R_{\mbox{\scriptsize {C}}}}
\def\vh{v_{\mbox{\scriptsize {H}}}}
\def\vtpar{v^\parallel_{\mbox{\scriptsize {TA}}}}
\def\vtper{v^\perp_{\mbox{\scriptsize {TA}}}}

\def\lb{\lambda_B}

\bibpunct{[}{]}{,\!}{n}{,}{,} 
\begin{document}
\title{Resonant Phonon Scattering in Quantum Hall Systems Driven by dc Electric Fields}

\author{W. Zhang}
\affiliation{School of Physics and Astronomy, University of Minnesota, Minneapolis, Minnesota 55455, USA} 
\author{M.\,A. Zudov}
\email[Corresponding author: ]{zudov@physics.umn.edu}
\affiliation{School of Physics and Astronomy, University of Minnesota, Minneapolis, Minnesota 55455, USA} 
\author{L.\,N. Pfeiffer}
\author{K.\,W. West}
\affiliation{Bell Labs, Alcatel-Lucent, Murray Hill, New Jersey 07974, USA}

\begin{abstract}
Using dc excitation to spatially tilt Landau levels, we study resonant acoustic phonon scattering in two-dimensional electron systems.
We observe that dc electric field strongly modifies phonon resonances, transforming resistance maxima into minima and back into maxima. 
Further, phonon resonances are enhanced dramatically in the non-linear dc response and can be detected even at low temperatures.
Most of our observations can be explained in terms of dc-induced (de)tuning of the resonant acoustic phonon scattering and its interplay with intra-Landau level impurity scattering. 
Finally, we observe a dc-induced zero-differential resistance state and a resistance maximum which occurs when the electron drift velocity approaches the speed of sound.
\end{abstract} 
\maketitle

Magneto-phonon resonance in two-dimensional electron systems (2DES) was studied long time ago \citep{gurevich:1961,tsui:1980}.
Experiments revealed that resistance oscillates with $\olo/\oc$, where $\olo$ is the frequency of longitudinal {\em optical} phonon and $\oc=eB/m^*$ ($m^*$ is the effective mass) is the cyclotron frequency.  
This effect requires high temperatures $T\sim 10^2$ K and high magnetic fields $B \sim 10^2$ kG. 

Over the past few years other types of resistance oscillations were discovered at much lower $T\sim 1$ K and $B\sim 1$ kG. 
Among these are microwave (ac)-induced resistance oscillations (MIRO) \citep{zudov:2001a,ye:2001}, Hall field (dc)-induced resistance oscillations \citep{yang:2002a,bykov:2005b,zhang:2007a} (HIRO), and oscillations due to resonant interaction with {\em acoustic} phonons which, for brevity, will be termed phonon-induced resistance oscillations (PIRO) \citep{zudov:2001b,yang:2002b,bykov:2005c}.
Remarkably, ac and dc excitations can lead to zero-resistance \citep{mani:2002, zudov:2003} and zero-differential resistance \cite{zhang:2007c, bykov:2007} states.
The majority of the theoretical work \citep{theory} targeted ac-induced effects but unresolved questions remain \citep{smet:2005,yang:2006}.

Stepping from inter-Landau level (LL) transitions, all induced resistance oscillations (IRO) are governed by the ratio of some relevant energy to $\hbar \oc$.
In MIRO, the relevant energy is the photon energy, $\hbar \omega$ and the resistance peaks occur at $\eac\equiv\omega/\oc \simeq  n - \pac$ ($n =1,2,...$ and $\pac \lesssim 1/4$) \citep{zudov:2004,studenikin:2005}.
In HIRO, observed in differential resistance, the peaks are found at $\edc\equiv \oh/\oc \simeq n =1,2,3,...$ \citep{zhang:2007a,lei:2007,vavilov:2007}. 
Here, $\hbar\oh \simeq e E (2\rc)$ is the energy associated with the Hall voltage drop across the cyclotron diameter ($E$ is the Hall field, $\rc$ is the cyclotron radius) \citep{yang:2002a}.
Similarly, PIRO peaks were associated with $\eph\equiv\os/\oc\simeq n = 1,2,3,...$, where $\os \simeq 2 k_F s$ ($k_F$ is the Fermi wave number, $s$ is the sound velocity) \citep{zudov:2001b}. 
Since all relevant energies are $B$-independent, these IRO are periodic in $1/B$ but their relative amplitudes differ greatly (MIRO:\,$\lesssim$$10^3$\%, HIRO:\,$\lesssim$$10^2$\%, PIRO:\,$\lesssim$$10^1$\%).
Further, all IRO extend to magnetic fields an order of magnitude lower than the onset of the Shubnikov-de Haas oscillations.
Finally, MIRO and HIRO are best observed at $T\sim 1$ K, but PIRO rely on thermal excitation of $2k_F$ phonons and require $T \sim 10$ K.

\begin{figure}[t]
\includegraphics{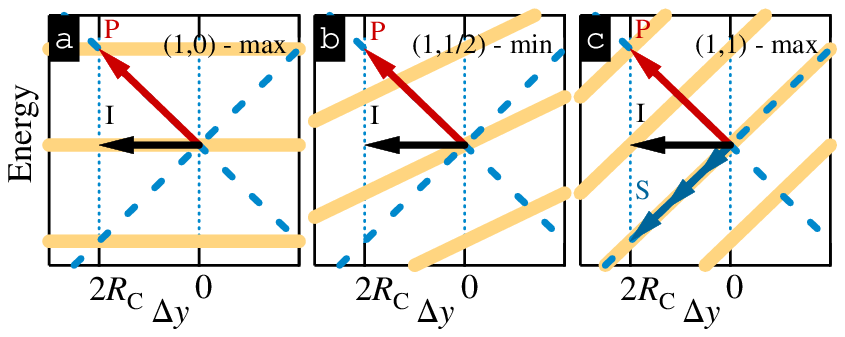}
\caption{[color online]
Solid lines are LLs at $\eph = 1$ and (a)\,$\edc=0$, (b)\,$\edc=1/2$, and (c)\,$\edc=1$ as functions of $\Delta y$.
Dashed lines represent phonon dispersion cone, $\hbar\os = \pm \hbar s \Delta y/\lb^2$.
Arrows show $2k_F$ transitions due to acoustic phonon (P) and impurity (I) scattering. 
At $\edc=\eph$\,(c), intra-LL scattering (S) accompanied by phonon emission becomes possible.
}
\label{lls}
\end{figure}

{\em Resonant} interaction of electrons with {\em acoustic} phonons is made possible in high LLs by virtue of a selection rule which favors electron backscattering, equivalent to a jump of the electron guiding center $\Delta y$ by a maximum distance, $\Delta y \simeq  2\rc$ \citep{zudov:2001b,yang:2002a}.
This largest jump is accompanied by the largest momentum transfer $\Delta q_x \equiv \Delta y/\lb^2 \simeq  2\rc/\lb^2 = 2k_F$ ($\lb = \sqrt{\hbar/eB}$ is the magnetic length), which can be supplied by an acoustic phonon.
The $2k_F$ acoustic phonon has a well defined energy $\hbar\os \simeq 2\hbar k_F s$ and thus can be resonantly absorbed by an electron jumping to a higher Landau level.
As a result, PIRO relate to {\em indirect} inter-LL transitions \citep{zudov:2001b} such as that shown by arrow (P) in Fig.\,\ref{lls}(a).

PIRO were originally explained in terms of {\em interface} phonons \citep{zudov:2001b,ponomarev:2001}.
In addition to the dominant mode with $s\simeq 2.9$ km/s, Fourier analysis suggested contribution from another mode with $s\simeq 4.4$ km/s. 
Detailed temperature dependence studies \citep{yang:2002b} confirmed these results.
In contrast, phonon-induced oscillations in magneto-thermopower were explained by a single {\em bulk} acoustic mode \citep{zhang:2004}. 
More recent magnetotransport studies \citep{bykov:2005c} have also reported a single mode, but with $s=5.9$ km/s.
Obviously, PIRO remain poorly understood and require further inverstigation. 
Of particular interest is the exact resonant condition, relative contribution of different modes and how it is affected by temperature and sample parameters.
Due to the relative weakness of PIRO and the lack of a systematic theory, these issues await future studies.
Here we concentrate on the effect of the dc electric field assuming a single phonon mode and traditionally associating the PIRO peaks with $\eph\simeq n$.

Experimentally, introducing additional parameters, e.g. second microwave frequency \citep{zudov:2006a, zudov:2006b}, in-plane $B$ \citep{mani:2005,yang:2006}, or dc excitation \citep{zhang:2007c}, proved to be powerful in studies of MIRO/ZRS.
Since dc excitation spatially tilts LLs due to the Hall effect but does not affect phonon dispersion, it might be used to tune the phonon resonances and to study the interplay between resonant phonon and impurity scattering.
For instance, as illustrated in Fig.\,\ref{lls}(b), dc current corresponding to $\edc = 1/2$ should take the phonon transition (P) shown in Fig.\,\ref{lls}(a) out of resonance.
Similarly, further tilting of LLs to $\edc \simeq 1$ should bring the phonon transition (P) back into the resonance, as illustrated in Fig.\,\ref{lls}(c).

Another question is related to dc excitations capable to accelerate electrons to the speed of sound.
Such a scenario was considered theoretically in relation to the ZRS breakdown \citep{ryzhii:2003c} but experiments have shown that ZRS disappear at much smaller currents \citep{zhang:2007c}.
At the crossover to the supersonic regime, LLs become parallel to one of the phonon branches [cf.\,Fig.\,\ref{lls}(c)] and intra-LL transitions (S) accompanied by phonon emission become possible.
Note that this crossover is expected to occur at $\edc = \eph$ but neither $\edc$ nor $\eph$ have to be integers. 

In this Letter we study magnetotransport in 2DES in the regime of elevated temperatures and dc currents $I$ up to 240 $\mu$A, which ensures that the drift velocity $\vh=I/en_ew$ exceeds the sound velocity.
While similar results were obtained from several different 2DES all the data presented here are from a Hall bar sample (width $w\!=\!50$ $\mu$m) fabricated from a GaAs/Al$_{0.24}$Ga$_{0.76}$As quantum well with
the low-temperature electron mobility $\mu \simeq  4.4 \times 10^6$ cm$^2$/Vs and the density $n_e = 4.8 \times 10^{11}$ cm$^{-2}$.
The differential resistivity $r\equiv dV/dI$ was measured with a lock-in amplifier supplying an ac (a few Hz) current of 0.1\,$\mu$A.

\begin{figure}[t]
\includegraphics{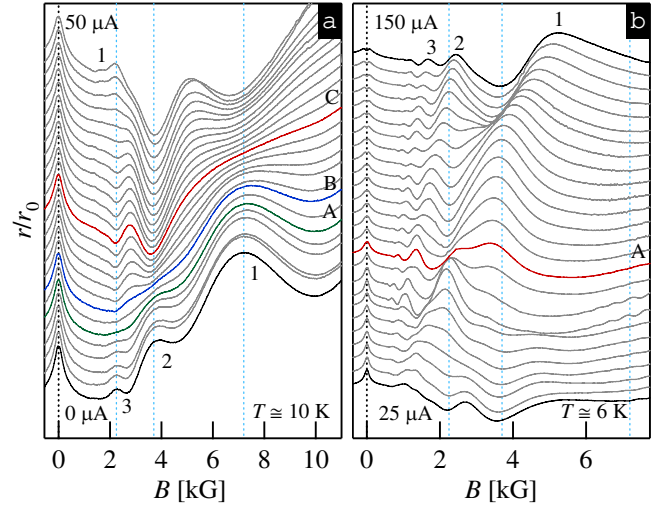}
\caption{[color online]
(a)\,[(b)] $r/r_0$ vs $B$ at $I$ from 0 to 50\,$\mu$A with step of 2\,$\mu$A ($T\simeq  10$ K) [from 25\,$\mu$A to 150\,$\mu$A with step of 5\,$\mu$A ($T\simeq 6$ K)].
Traces are vertically offset in increments of 2\%\,[15\%]. 
Vertical lines correspond to the PIRO peaks as marked at $I=0$. Numbers at the top traces mark HIRO peaks.
}
\label{bsweep}
\end{figure}

In Fig.\,\ref{bsweep}(a) we present $r/r_0$ ($r_0$ is the zero-$B$ value of $r$) as a function of $B$ at $I$ (in $\mu$A throughout the paper) from 0 to 50 in steps of 2 at $T\simeq  10$ K. 
The data at $I=0$ (cf., bottom trace) show a series of PIRO peaks, marked by vertical lines, which roughly correspond to $\eph=1,2,3$ with $s\simeq 5.8$\,km/s.
We note that Shubnikov-de Haas oscillations are not yet observable in this $B$ range and that the peak at $\eph=1$ is strongly suppressed at lower temperatures possibly indicating dominance of slower modes as in earlier studies \citep{zudov:2001b}.

We now examine how PIRO are affected by dc excitation.
Since HIRO remain below $B\simeq 1.7$ kG [cf.\,``1'' at the top ($I=50$) trace in Fig.\,\ref{bsweep}(a)] and PIRO start at higher $B$, no direct mixing of the oscillations is possible.
Nevertheless, we observe that PIRO undergo dramatic changes.
The third PIRO peak disappears at $I\simeq 10$ (cf.\,trace A), then reappears as a minimum (cf.\,traces B,C), and eventually reverts back to a peak (cf.\,top trace). 
Second and first PIRO peaks vanish at $I\simeq  14$ (cf.\,trace B) and $I\simeq  28$ (cf.\,trace C), respectively, and become minima at higher $I$ (cf.\,top trace).
These observations suggest strong coupling of dc excitation and electron-phonon scattering, as otherwise one expects only {\em overall} suppression of resistance without evolution of maxima into minima and vice versa \citep{zhang:2007a, zhang:2007b}.

In Fig.\,\ref{bsweep}(b) we continue with the data at higher $I$ (up to 150, in steps of 5) and lower temperature $T\simeq 6$ K, where PIRO and HIRO coesxist.
The results show that while HIRO appear to dominate in this regime, their waveform is strongly influenced by phonon scattering.
Being a deep minimum at $I=25$, $\eph\simeq 2$ again becomes a peak at $I\simeq  75$, (cf.\,trace A) and then merges with $\edc \simeq 1$ maximum.
This occurs when both $\eph$ and $\edc$ are close to integers and therefore resonant conditions for both phonon (P) and impurity (I) scattering are satisfied, similar to the situation illustrated in Fig.\ref{lls}(c).
It turns out that the PIRO maxima are converted into the minima when $\edc \simeq 1/2$ and then back to the maxima when $\edc \simeq 1$.
This is consistent with dc-induced detuning from the magneto-phonon resonances as depicted in Fig.\,\ref{lls}(b) and (c).
We therefore conclude that at fixed $\eph$ oscillations are periodic in $\edc$. 

\begin{figure}[t]
\includegraphics{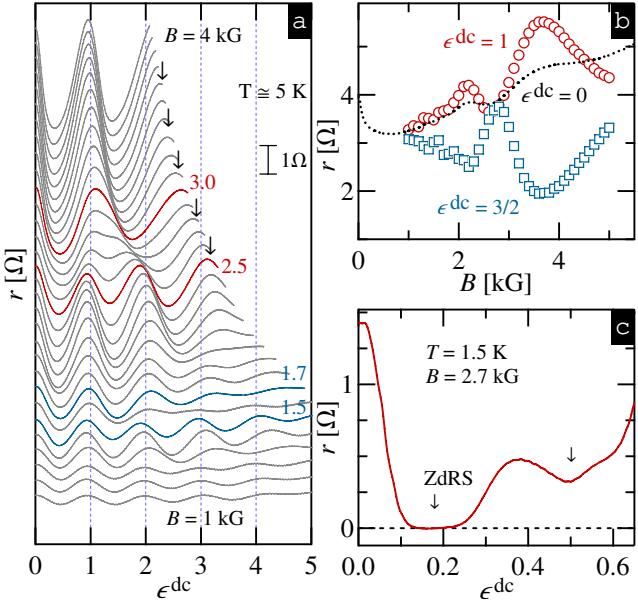}
\caption{[color online]
(a) $r$ vs $\edc$ at $B$ from 1 kG to 4 kG in 0.1 kG steps ($T\simeq  5$ K). 
Traces are vertically offset by 0.5 $\Omega$. 
(b) $r$ vs $B$ at $\edc=0$\,(dotted line), $\edc=1$\,(circles), and $\edc=3/2$ (squares) at $T\simeq  5$ K. 
(c) $r$ vs $\edc$ at $B=2.7$ kG and $T\simeq 1.5$\,K. 
}
\label{isweep}
\end{figure}

As an alternative way to study the effect of dc excitation we perform current sweeps up to $I=240$ at fixed $B$ from 1 to 4 kG (in 0.1 kG steps) and at $T \simeq  5$ K. 
The results are presented in Fig.\,\ref{isweep}(a) showing $r$ as a function of $\edc$.
The most striking feature is the transformation of the $\edc \simeq 2$ peak into a minimum taking place between 2.5 and 3.0 kG.
Similar behavior is also seen at other $B$, e.g., between 1.5 and 1.7 kG, where the minimum emerges near $\edc \simeq 3$.
These ranges of $B$ correspond to $3\gtrsim \eph \gtrsim 2.5$ and $5\gtrsim \eph \gtrsim 4.5$, respectively.
At lower $B$, $\eph$ and $\edc$ are both integers, and both impurity and phonon scattering are resonantly enhanced [cf.\,Fig.\,\ref{lls}(c)].
In contrast, at higher $B$,  $\eph$ becomes half-integer and a minimum is observed.
Another interesting observation is the peak in resistance [cf.\,$\downarrow$] which gradually moves toward lower $\edc$ with increasing $B$.
As we will see, this peak marks the crossover to the supersonic regime.  

We now examine PIRO at fixed integer and half-integer values of $\edc$.
In Fig.\,\ref{isweep}(b) we present $r$ as a function of $B$ at $\edc \simeq 1$ (circles) and at $\edc \simeq 3/2$ (squares).
Both data sets show prominent oscillation with the same period as the zero-bias trace (dotted line) but with a greatly enhanced amplitude. 
However, the $\edc \simeq 1$ data are in-phase with the zero-bias trace, while the $\edc \simeq 3/2$ data are out-of-phase.
This leads to the conclusion that at fixed $\edc$, oscillations are periodic in $\eph$.
The fact that PIRO translate to integer $\edc$ is not surprising since both phenomena rely on $2k_F$ momentum transfer.
Note that contrary to ac/dc-induced oscillations \citep{zhang:2007b} phonon resonance remains ``phased-locked'' to the impurity resonances since {\em both} require the same momentum transfer.
This is illustrated in Fig. \ref{lls} where (P) and (I) processes are either {\em both} out-of-resonance [panel (b)] or {\em both} in-resonance [panel (c)].
Remarkably, as a result of these ``double'' resonances PIRO are amplified and can be detected even at low $T\simeq 1.5$ K, even though they are not observed at $I=0$.
However, what really distinguishes the $T\simeq 1.5$ K data is the formation of the dc-induced zero-differential resistance states, an example of which is shown in Fig.\,\ref{isweep}(c). 
While quite striking, these states appear unrelated to phonon scattering and thus fall outside the scope of this paper.

We can summarize our observations as follows. 
If $\eph$ or $\edc$ are integers or half integers, then the resistance maxima (minima) occur close to integer (half-integer) values of a  parameter $\epsilon = \eph + \edc$:
\begin{equation}
\epsilon_{+}\simeq n,\,\,\epsilon_{-}\simeq n+1/2,\,\,n=0,1,2,...
\label{eq}
\end{equation}
Note that Eq.\,\ref{eq} implies that the PIRO peaks should gradually move towards higher B with increasing I, which is not readily observed in Fig.\,\ref{bsweep}(a). 
Instead, in this range of dc currents PIRO peaks appear to evolve into minima without obvious change in the $B$ position, much the same as the MIRO peaks \citep{zhang:2007c}.
We recall that the MIRO peaks under dc excitation roughly conform to $\eac+\edc \simeq n$ but show deviations at $\eac$ or $\edc$ away from integer and half-integer values.
These deviations are important as they indicate the limitations of the applicability of Eq.\,\ref{eq} and call for a systematic theoretical treatment.

\begin{figure}[t]
\includegraphics{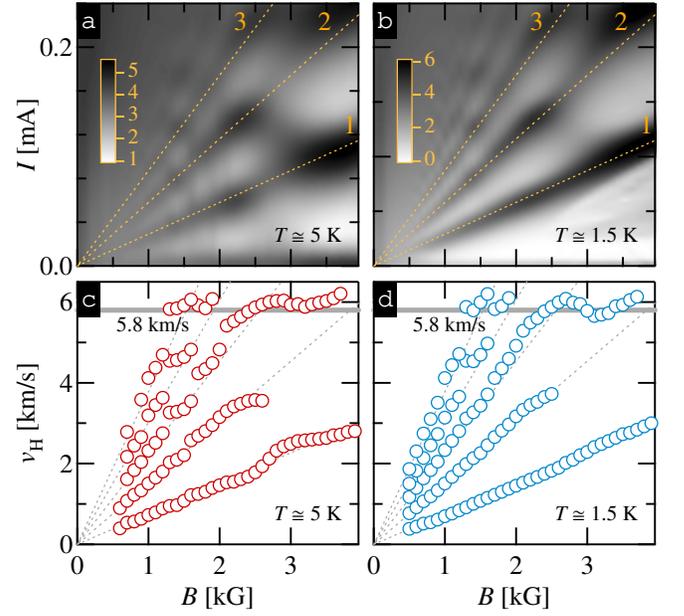}
\caption{[color online]
(a) [(b)] $r$ in the ($B$, $I$) plane at $T\simeq  5$ K [$T\simeq  1.5$ K]. 
Greyscale is in $\Omega$.
(c) [(d)] $r$ maxima in ($B,\vh$)-plane at $T\simeq  5$ K [$T\simeq  1.5$ K]. 
Horizontal line is at 5.8 km/s.
}
\label{im}
\end{figure}

Using the data such as shown in Fig.\,\ref{isweep}(a) we construct greyscale plots of $r$ in the ($B$, $I$)-plane which are shown in Fig.\,\ref{im}(a)\,[(b)] for $T\simeq 5\,[1.5]$ K.
If one neglects the effect of phonons, the maxima of $r$ are expected to fall onto a fan diagram in accordance with $\edc=n$ which is represented by dotted lines emanating from the origin.
Despite general agreement, we readily observe periodic modulation along the fan lines which is just the signal presented in Fig.\,\ref{isweep}(b).
Alternatively, by moving in the vertical direction, one recovers the oscillations presented in Fig.\,\ref{isweep}(a).
As a result, the overall view reveals a matrix of strong peaks due to ``double'' resonances close to integer $\eph$ and $\edc$.

We now extract the peak positions and present the results in Fig.\,\ref{im}(c)\,[(d)] for $T\simeq 5\,[1.5]$ K, while converting $I$ to the drift velocity $\vh$.
The most striking observation is the deviation from the fan observed where the peaks roughly follow the horizontal line near $\vh \simeq 5.8$ km/s.
Since this is the value extracted from the zero-bias data in Fig.\,\ref{bsweep}(a), these peaks mark a crossover to the supersonic regime at $\vh \simeq  s$ and can be qualitatively understood as follows. 
Once the Hall electric field reaches the critical value, $E_s=esB$ (or, equivalently, when $\edc = \eph$), Landau levels become ``parallel'' to the phonon dispersion.
This opens a scattering channel due to intra-LL transitions accompanied by phonon emission as illustrated in Fig.\,\ref{lls}(c).
Note that in such processes the energy and momentum conservation can be satisfied at {\em any} $B$ (and at any $q$). 

On the other hand, the origin of the deviations at lower $\vh$ is not so clear. 
While these can arise from the interplay between phonon and impurity scattering, one cannot rule out the possibility of crossovers to the supersonic regime corresponding to slower acoustic modes.
In particular, some prominent deviations occur near 3.5 km/s and 2.6 km/s, which are close to velocities of transverse acoustic modes $\vtpar = 3.4$ km/s and $\vtper = 2.5$ km/s.
We speculate that since phonon scattering is enhanced at finite values of dc excitation, other modes, not dominant in linear transport,  might become relevant.

In summary, we have studied phonon-induced resistance oscillations of a high-quality 2DES under dc excitation.
First, we have shown that dc excitation causes evolution of the PIRO maxima into minima and back.
This evolution can be understood in terms of dc-induced (de)tuning of phonon resonances as illustrated in Fig.\,\ref{lls}.
Second, we have found that PIRO are dramatically enhanced by a strong dc field and can be detected even at low temperatures where linear resistivity is usually {\em not} sensitive to phonons.
This enhancement can be ascribed to ``double'' resonances occurring because {\em both} phonon and impurity scattering remain in-phase. 
Third, we have observed a pronounced resistance maximum which appears when the drift velocity approaches the speed of sound.
We explain this peak in terms of intra-LL transitions accompanied by phonon emission.
Finally, we have detected differential zero-resistance states which are induced by pure dc-excitation at low temperatures.
On the other hand, detailed understanding of the PIRO evolution with increasing current, dc-induced enhancement, and the relevance of other phonon modes remain a subject of further theoretical and experimental studies.

We thank I. Dmitriev, A. Kamenev, B. Shklovskii, and M. Vavilov for useful discussions.
This work was supported by NSF Grant No. DMR-0548014.
\vspace{-0.10in}

\begin{thebibliography}{29}
\expandafter\ifx\csname natexlab\endcsname\relax\def\natexlab#1{#1}\fi
\expandafter\ifx\csname bibnamefont\endcsname\relax
  \def\bibnamefont#1{#1}\fi
\expandafter\ifx\csname bibfnamefont\endcsname\relax
  \def\bibfnamefont#1{#1}\fi
\expandafter\ifx\csname citenamefont\endcsname\relax
  \def\citenamefont#1{#1}\fi
\expandafter\ifx\csname url\endcsname\relax
  \def\url#1{\texttt{#1}}\fi
\expandafter\ifx\csname urlprefix\endcsname\relax\def\urlprefix{URL }\fi
\providecommand{\bibinfo}[2]{#2}
\providecommand{\eprint}[2][]{\url{#2}}

\bibitem[{\citenamefont{Gurevich and Firsov}(1961)}]{gurevich:1961}
\bibinfo{author}{\bibfnamefont{V.~L.} \bibnamefont{Gurevich}} \bibnamefont{and}
  \bibinfo{author}{\bibfnamefont{Y.}~\bibnamefont{Firsov}},
  \bibinfo{journal}{Sov. Phys. JETP} \textbf{\bibinfo{volume}{13}},
  \bibinfo{pages}{137} (\bibinfo{year}{1961}).

\bibitem[{\citenamefont{Tsui et~al.}(1980)\citenamefont{Tsui, Englert, Cho, and
  Gossard}}]{tsui:1980}
\bibinfo{author}{\bibfnamefont{D.~C.} \bibnamefont{Tsui}},
  \bibinfo{author}{\bibfnamefont{T.}~\bibnamefont{Englert}},
  \bibinfo{author}{\bibfnamefont{A.~Y.} \bibnamefont{Cho}}, \bibnamefont{and}
  \bibinfo{author}{\bibfnamefont{A.~C.} \bibnamefont{Gossard}},
  \bibinfo{journal}{Phys. Rev. Lett.} \textbf{\bibinfo{volume}{44}},
  \bibinfo{pages}{341} (\bibinfo{year}{1980}).

\bibitem[{\citenamefont{Zudov et~al.}(2001{\natexlab{a}})\citenamefont{Zudov,
  Du, Simmons, and Reno}}]{zudov:2001a}
  \bibinfo{author}{\bibfnamefont{M.~A.} \bibnamefont{Zudov}},
  \bibinfo{author}{\bibfnamefont{R.~R.} \bibnamefont{Du}},
  \bibinfo{author}{\bibfnamefont{J.~A.} \bibnamefont{Simmons}},
  \bibnamefont{and} \bibinfo{author}{\bibfnamefont{J.~L.} \bibnamefont{Reno}},
  \bibinfo{journal}{Phys. Rev. B} \textbf{\bibinfo{volume}{64}},
  \bibinfo{eid}{201311(R)} (\bibinfo{year}{2001}{\natexlab{a}}).

\bibitem[{\citenamefont{Ye et~al.}(2001)\citenamefont{Ye, Engel, Tsui, Simmons,
  Wendt, Vawter, and Reno}}]{ye:2001}
\bibinfo{author}{\bibfnamefont{P.~D.} \bibnamefont{Ye}},
\bibinfo{author}{\bibfnamefont{L.~W.} \bibnamefont{Engel}},
\bibinfo{author}{\bibfnamefont{D.~C.} \bibnamefont{Tsui}},
\bibinfo{author}{\bibfnamefont{J.~A.} \bibnamefont{Simmons}},
\bibinfo{author}{\bibfnamefont{J.~R.} \bibnamefont{Wendt}},
\bibinfo{author}{\bibfnamefont{G.~A.} \bibnamefont{Vawter}},
\bibnamefont{and} \bibinfo{author}{\bibfnamefont{J.~L.} \bibnamefont{Reno}},
  \bibinfo{journal}{Appl. Phys. Lett.} \textbf{\bibinfo{volume}{79}},
  \bibinfo{eid}{2193} (\bibinfo{year}{2001}).

\bibitem[{\citenamefont{Yang et~al.}(2002{\natexlab{a}})\citenamefont{Yang,
  Zhang, Du, Simmons, and Reno}}]{yang:2002a}
\bibinfo{author}{\bibfnamefont{C.~L.} \bibnamefont{Yang}},
  \bibinfo{author}{\bibfnamefont{J.}~\bibnamefont{Zhang}},
  \bibinfo{author}{\bibfnamefont{R.~R.} \bibnamefont{Du}},
  \bibinfo{author}{\bibfnamefont{J.~A.} \bibnamefont{Simmons}},
  \bibnamefont{and} \bibinfo{author}{\bibfnamefont{J.~L.} \bibnamefont{Reno}},
  \bibinfo{journal}{Phys. Rev. Lett.} \textbf{\bibinfo{volume}{89}},
  \bibinfo{eid}{076801} (\bibinfo{year}{2002}{\natexlab{a}}).

\bibitem[{\citenamefont{Bykov et~al.}(2005{\natexlab{a}})\citenamefont{Bykov,
  Zhang, Vitkalov, Kalagin, and Bakarov}}]{bykov:2005b}
\bibinfo{author}{\bibfnamefont{A.~A.} \bibnamefont{Bykov}},
  \bibinfo{author}{\bibfnamefont{J.~Q.}~\bibnamefont{Zhang}},
  \bibinfo{author}{\bibfnamefont{S.}~\bibnamefont{Vitkalov}},
  \bibinfo{author}{\bibfnamefont{A.~K.} \bibnamefont{Kalagin}},
  \bibnamefont{and} \bibinfo{author}{\bibfnamefont{A.~K.}
  \bibnamefont{Bakarov}}, \bibinfo{journal}{Phys. Rev. B}
  \textbf{\bibinfo{volume}{72}}, \bibinfo{eid}{245307}
  (\bibinfo{year}{2005}{\natexlab{a}}).

\bibitem[{\citenamefont{Zhang et~al.}(2007{\natexlab{a}})\citenamefont{Zhang,
  Chiang, Zudov, Pfeiffer, and West}}]{zhang:2007a}
\bibinfo{author}{\bibfnamefont{W.}~\bibnamefont{Zhang}},
  \bibinfo{author}{\bibfnamefont{H.-S.} \bibnamefont{Chiang}},
  \bibinfo{author}{\bibfnamefont{M.~A.} \bibnamefont{Zudov}},
  \bibinfo{author}{\bibfnamefont{L.~N.} \bibnamefont{Pfeiffer}},
  \bibnamefont{and} \bibinfo{author}{\bibfnamefont{K.~W.} \bibnamefont{West}},
  \bibinfo{journal}{Phys. Rev. B} \textbf{\bibinfo{volume}{75}},
  \bibinfo{eid}{041304(R)} (\bibinfo{year}{2007}{\natexlab{a}}).

\bibitem[{\citenamefont{Zudov et~al.}(2001{\natexlab{b}})\citenamefont{Zudov,
  Ponomarev, Efros, Du, Simmons, and Reno}}]{zudov:2001b}
\bibinfo{author}{\bibfnamefont{M.~A.} \bibnamefont{Zudov}},
  \bibinfo{author}{\bibfnamefont{I.~V.} \bibnamefont{Ponomarev}},
  \bibinfo{author}{\bibfnamefont{A.~L.} \bibnamefont{Efros}},
  \bibinfo{author}{\bibfnamefont{R.~R.} \bibnamefont{Du}},
  \bibinfo{author}{\bibfnamefont{J.~A.} \bibnamefont{Simmons}},
  \bibnamefont{and} \bibinfo{author}{\bibfnamefont{J.~L.} \bibnamefont{Reno}},
  \bibinfo{journal}{Phys. Rev. Lett.} \textbf{\bibinfo{volume}{86}},
  \bibinfo{pages}{3614} (\bibinfo{year}{2001}{\natexlab{b}}).

\bibitem[{\citenamefont{Yang et~al.}(2002{\natexlab{b}})\citenamefont{Yang,
  Zudov, Zhang, Du, Simmons, and Reno}}]{yang:2002b}
\bibinfo{author}{\bibfnamefont{C.~L.} \bibnamefont{Yang}},
  \bibinfo{author}{\bibfnamefont{M.~A.} \bibnamefont{Zudov}},
  \bibinfo{author}{\bibfnamefont{J.}~\bibnamefont{Zhang}},
  \bibinfo{author}{\bibfnamefont{R.~R.} \bibnamefont{Du}},
  \bibinfo{author}{\bibfnamefont{J.~A.} \bibnamefont{Simmons}},
  \bibnamefont{and} \bibinfo{author}{\bibfnamefont{J.~L.} \bibnamefont{Reno}},
  \bibinfo{journal}{Physica E} \textbf{\bibinfo{volume}{12}},
  \bibinfo{pages}{443} (\bibinfo{year}{2002}{\natexlab{b}}).

\bibitem[{\citenamefont{Bykov et~al.}(2005{\natexlab{b}})\citenamefont{Bykov,
  Kalagin, and Bakarov}}]{bykov:2005c}
\bibinfo{author}{\bibfnamefont{A.~A.} \bibnamefont{Bykov}},
  \bibinfo{author}{\bibfnamefont{A.~K.} \bibnamefont{Kalagin}},
  \bibnamefont{and} \bibinfo{author}{\bibfnamefont{A.~K.}
  \bibnamefont{Bakarov}}, \bibinfo{journal}{JETP Lett.}
  \textbf{\bibinfo{volume}{81}}, \bibinfo{pages}{523}
  (\bibinfo{year}{2005}{\natexlab{b}}).

\bibitem[{\citenamefont{Mani et~al.}(2002)\citenamefont{Mani, Smet, von
  Klitzing, Narayanamurti, Johnson, and Umansky}}]{mani:2002}
\bibinfo{author}{\bibfnamefont{R.~G.} \bibnamefont{Mani}},
  \bibinfo{author}{\bibfnamefont{J.~H.} \bibnamefont{Smet}},
  \bibinfo{author}{\bibfnamefont{K.}~\bibnamefont{von Klitzing}},
  \bibinfo{author}{\bibfnamefont{V.}~\bibnamefont{Narayanamurti}},
  \bibinfo{author}{\bibfnamefont{W.~B.} \bibnamefont{Johnson}},
  \bibnamefont{and} \bibinfo{author}{\bibfnamefont{V.}~\bibnamefont{Umansky}},
  \bibinfo{journal}{Nature} \textbf{\bibinfo{volume}{420}},
  \bibinfo{pages}{646} (\bibinfo{year}{2002}).

\bibitem[{\citenamefont{Zudov et~al.}(2003)\citenamefont{Zudov, Du, Pfeiffer,
  and West}}]{zudov:2003}
\bibinfo{author}{\bibfnamefont{M.~A.} \bibnamefont{Zudov}},
  \bibinfo{author}{\bibfnamefont{R.~R.} \bibnamefont{Du}},
  \bibinfo{author}{\bibfnamefont{L.~N.} \bibnamefont{Pfeiffer}},
  \bibnamefont{and} \bibinfo{author}{\bibfnamefont{K.~W.} \bibnamefont{West}},
  \bibinfo{journal}{Phys. Rev. Lett.} \textbf{\bibinfo{volume}{90}},
  \bibinfo{eid}{046807} (\bibinfo{year}{2003}).

\bibitem[{\citenamefont{Zhang et~al.}(2007{\natexlab{b}})\citenamefont{Zhang,
  Zudov, Pfeiffer, and West}}]{zhang:2007c}
\bibinfo{author}{\bibfnamefont{W.}~\bibnamefont{Zhang}},
\bibinfo{author}{\bibfnamefont{M.~A.} \bibnamefont{Zudov}},
\bibinfo{author}{\bibfnamefont{L.~N.} \bibnamefont{Pfeiffer}},
\bibnamefont{and} \bibinfo{author}{\bibfnamefont{K.~W.} \bibnamefont{West}},
  \bibinfo{journal}{Phys. Rev. Lett.} \textbf{\bibinfo{volume}{98}},
  \bibinfo{pages}{106804} (\bibinfo{year}{2007}{\natexlab{b}}); 
  Physica E, doi:10.1016/j.physe.2007.08.066 (2007).

\bibitem[{\citenamefont{Bykov et~al.}(unpublished)\citenamefont{Bykov, Zhang,
  Vitkalov, Kalagin, and Bakarov}}]{bykov:2007}
\bibinfo{author}{\bibfnamefont{A.~A.} \bibnamefont{Bykov}},
  \bibinfo{author}{\bibfnamefont{J.-Q.} \bibnamefont{Zhang}},
  \bibinfo{author}{\bibfnamefont{S.}~\bibnamefont{Vitkalov}},
  \bibinfo{author}{\bibfnamefont{A.~K.} \bibnamefont{Kalagin}},
  \bibnamefont{and} \bibinfo{author}{\bibfnamefont{A.~K.}
  \bibnamefont{Bakarov}}, \bibinfo{journal}{Phys. Rev. Lett.} \textbf{\bibinfo{volume}{99}},
  \bibinfo{pages}{116801} (\bibinfo{year}{2007}).

\bibitem[{the()}]{theory}
\bibinfo{note}{V.~I.~Ryzhii, Sov.\ Phys.\ Solid State {\bf 11}, 2078 (1970);
  A.~V.~Andreev {\em et al.}, Phys.\ Rev.\ Lett.\ {\bf 91}, 056803 (2003);
  J.~Shi and X.~C.~Xie, {\em ibid.} {\bf 91}, 086801 (2003);  
  A.~C.~Durst {\em et al.}, {\em ibid.} {\bf 91}, 086803 (2003); 
  I.~A.~Dmitriev {\em et al.}, {\em ibid.} {\bf 91}, 226802 (2003); 
  X.~L.~Lei and S.~Y.~Liu, {\em ibid.} {\bf 91}, 226805 (2003);
  J. Inarrea and G. Platero, {\em ibid.} {\bf 94}, 016806 (2005); 
  A.~Auerbach {\em et al.}, {\em ibid.} {\bf 94}, 196801 (2005); 
  I.~A.~Dmitriev {\em et al.}, {\em ibid.} {\bf 99}, to appear (2007); 
  M.~G.~Vavilov and I.~L.~Aleiner, Phys.\ Rev.\ B {\bf 69}, 035303 (2004); 
  I.~A.~Dmitriev {\em et al.}, {\em ibid.} {\bf 70}, 165305 (2004); {\em ibid.} {\bf 71}, 115316 (2005); {\em ibid.} {\bf 75}, 245320 (2007) 
  (and references therein)
  }.

\bibitem[{\citenamefont{Smet et~al.}(2005)\citenamefont{Smet, Gorshunov, Jiang,
  Pfeiffer, West, Umansky, Dressel, Meisels, Kuchar, and von
  Klitzing}}]{smet:2005}
\bibinfo{author}{\bibfnamefont{J.~H.} \bibnamefont{Smet {\it et~al.}}},
  \bibinfo{journal}{Phys. Rev. Lett.} \textbf{\bibinfo{volume}{95}},
  \bibinfo{eid}{116804} (\bibinfo{year}{2005}).

\bibitem[{\citenamefont{Yang et~al.}(2006)\citenamefont{Yang, Du, Pfeiffer, and
  West}}]{yang:2006}
\bibinfo{author}{\bibfnamefont{C.~L.} \bibnamefont{Yang}},
  \bibinfo{author}{\bibfnamefont{R.~R.} \bibnamefont{Du}},
  \bibinfo{author}{\bibfnamefont{L.~N.} \bibnamefont{Pfeiffer}},
  \bibnamefont{and} \bibinfo{author}{\bibfnamefont{K.~W.} \bibnamefont{West}},
  \bibinfo{journal}{Phys. Rev. B} \textbf{\bibinfo{volume}{74}},
  \bibinfo{eid}{045315} (\bibinfo{year}{2006}).

\bibitem[{\citenamefont{Zudov}(2004)}]{zudov:2004}
\bibinfo{author}{\bibfnamefont{M.~A.} \bibnamefont{Zudov}},
  \bibinfo{journal}{Phys. Rev. B} \textbf{\bibinfo{volume}{69}},
  \bibinfo{eid}{041304(R)} (\bibinfo{year}{2004}).

\bibitem[{\citenamefont{Studenikin et~al.}(2005)\citenamefont{Studenikin,
  Potemski, Sachrajda, Hilke, Pfeiffer, and West}}]{studenikin:2005}
\bibinfo{author}{\bibfnamefont{S.~A.} \bibnamefont{Studenikin {\it et~al.}}},
  \bibinfo{journal}{Phys. Rev. B} \textbf{\bibinfo{volume}{71}},
  \bibinfo{eid}{245313} (\bibinfo{year}{2005}).

\bibitem[{\citenamefont{Lei}(2007)}]{lei:2007}
\bibinfo{author}{\bibfnamefont{X.~L.} \bibnamefont{Lei}},
  \bibinfo{journal}{Appl. Phys. Lett.} \textbf{\bibinfo{volume}{90}},
  \bibinfo{pages}{132119} (\bibinfo{year}{2007}).

\bibitem[{\citenamefont{Vavilov et~al.}(2007)\citenamefont{Vavilov,
  Aleiner, and Glazman}}]{vavilov:2007}
\bibinfo{author}{\bibfnamefont{M.~G.} \bibnamefont{Vavilov}},
  \bibinfo{author}{\bibfnamefont{I.~L.} \bibnamefont{Aleiner}},
  \bibnamefont{and} \bibinfo{author}{\bibfnamefont{L.~I.}
  \bibnamefont{Glazman}}, \bibinfo{journal}{Phys. Rev. B} \textbf{\bibinfo{volume}{76}},
  \bibinfo{eid}{115331} (\bibinfo{year}{2007}).

\bibitem[{\citenamefont{Ponomarev and Efros}(2001)}]{ponomarev:2001}
\bibinfo{author}{\bibfnamefont{I.~V.} \bibnamefont{Ponomarev}}
  \bibnamefont{and} 
  \bibinfo{author}{\bibfnamefont{A.~L.} \bibnamefont{Efros}},
  \bibinfo{journal}{Phys. Rev. B} \textbf{\bibinfo{volume}{63}},
  \bibinfo{eid}{165305} (\bibinfo{year}{2001}).

\bibitem[{\citenamefont{Zhang et~al.}(2004)\citenamefont{Zhang, Lyo, Du,
  Simmons, and Reno}}]{zhang:2004}
\bibinfo{author}{\bibfnamefont{J.}~\bibnamefont{Zhang}},
  \bibinfo{author}{\bibfnamefont{S.~K.} \bibnamefont{Lyo}},
  \bibinfo{author}{\bibfnamefont{R.~R.} \bibnamefont{Du}},
  \bibinfo{author}{\bibfnamefont{J.~A.} \bibnamefont{Simmons}},
  \bibnamefont{and} \bibinfo{author}{\bibfnamefont{J.~L.} \bibnamefont{Reno}},
  \bibinfo{journal}{Phys. Rev. Lett.} \textbf{\bibinfo{volume}{92}},
  \bibinfo{eid}{156802} (\bibinfo{year}{2004}).

\bibitem[{\citenamefont{Zudov et~al.}(2006{\natexlab{b}})\citenamefont{Zudov,
  Du, Pfeiffer, and West}}]{zudov:2006a}
\bibinfo{author}{\bibfnamefont{M.~A.} \bibnamefont{Zudov}},
  \bibinfo{author}{\bibfnamefont{R.~R.} \bibnamefont{Du}},
  \bibinfo{author}{\bibfnamefont{L.~N.} \bibnamefont{Pfeiffer}},
  \bibnamefont{and} \bibinfo{author}{\bibfnamefont{K.~W.} \bibnamefont{West}},
  \bibinfo{journal}{Phys. Rev. B} \textbf{\bibinfo{volume}{73}},
  \bibinfo{eid}{041303(R)} (\bibinfo{year}{2006}{\natexlab{b}}).

\bibitem[{\citenamefont{Zudov et~al.}(2006{\natexlab{a}})\citenamefont{Zudov,
  Du, Pfeiffer, and West}}]{zudov:2006b}
\bibinfo{author}{\bibfnamefont{M.~A.} \bibnamefont{Zudov}},
  \bibinfo{author}{\bibfnamefont{R.~R.} \bibnamefont{Du}},
  \bibinfo{author}{\bibfnamefont{L.~N.} \bibnamefont{Pfeiffer}},
  \bibnamefont{and} \bibinfo{author}{\bibfnamefont{K.~W.} \bibnamefont{West}},
  \bibinfo{journal}{Phys. Rev. Lett.} \textbf{\bibinfo{volume}{96}},
  \bibinfo{eid}{236804} (\bibinfo{year}{2006}{\natexlab{a}}).

\bibitem[{\citenamefont{Mani}(2005)}]{mani:2005}
\bibinfo{author}{\bibfnamefont{R.~G.} \bibnamefont{Mani}},
  \bibinfo{journal}{Phys. Rev. B} \textbf{\bibinfo{volume}{72}},
  \bibinfo{eid}{075327(R)} (\bibinfo{year}{2005}).
  


\bibitem[{\citenamefont{Ryzhii and Satou}(2003)}]{ryzhii:2003c}
\bibinfo{author}{\bibfnamefont{V.}~\bibnamefont{Ryzhii}} \bibnamefont{and}
  \bibinfo{author}{\bibfnamefont{A.}~\bibnamefont{Satou}}, \bibinfo{journal}{J.
  Phys. Soc. Jpn.} \textbf{\bibinfo{volume}{72}}, \bibinfo{pages}{2718}
  (\bibinfo{year}{2003}).

\bibitem[{\citenamefont{Zhang et~al.}(2007{\natexlab{c}})\citenamefont{Zhang,
  Vitkalov, Bykov, Kalagin, and Bakarov}}]{zhang:2007b}
\bibinfo{author}{\bibfnamefont{J.~Q.}~\bibnamefont{Zhang}},
  \bibinfo{author}{\bibfnamefont{S.}~\bibnamefont{Vitkalov}},
  \bibinfo{author}{\bibfnamefont{A.~A.} \bibnamefont{Bykov}},
  \bibinfo{author}{\bibfnamefont{A.~K.} \bibnamefont{Kalagin}},
  \bibnamefont{and} \bibinfo{author}{\bibfnamefont{A.~K.}
  \bibnamefont{Bakarov}}, \bibinfo{journal}{Phys. Rev. B}
  \textbf{\bibinfo{volume}{75}}, \bibinfo{pages}{081305(R)}
  (\bibinfo{year}{2007}{\natexlab{c}}).
  
\end{thebibliography}

\end{document}